# Characterization of casein and Poly-L-arginine Multilayer Films.

Lilianna Szyk – Warszyńska*, Katarzyna Kilan, Robert P. Socha
*Jerzy Haber Institute of Catalysis and Surface Chemistry Polish Academy of Sciences, Niezapominajek 8, 30-238 Kraków, Poland.*

**Abstract:** Thin films containing casein appear to be a promising material for coatings used in the medical area to promote biomineralization. α- and β-casein and poly-L-arginine multilayer films were formed by the layer-by layer technique and their thickness and mass were analyzed by ellipsometry and quartz crystal microbalance with dissipation monitoring (QCM-D). We investigated the effect of the type of casein used for the film formation and of the polyethyleneimine anchoring layer on the thickness and mass of adsorbed films. The analysis of the mass of films during their post-treatment with the solutions of various ionic strength and pH provided the information concerning films stability, while the XPS elemental analysis confirmed binding of calcium ions by the casein embedded in the multilayers.

**Highlights**

- Films containing casein together with polyaminoacid poly-L-arginine can be formed by the LbL technique.
- The multilayer films with α-casein were thicker than those with β-casein.
- Films remain stable in the neutral and weakly basic conditions while tend to disintegrate in pH < 4.
- At the conditions of high ionic strength films swell but that swelling is reversible.
- After exposure to CaCl2 solutions calcium ions can be bound by the casein embedded in the multilayers.

**Key words:** casein, layer-by-layer, protein adsorption, multilayers, ellipsometry, QCM-D, calcium binding

## 1. Introduction

Casein belongs to the heterogeneous group of phosphoproteins precipitated from raw skimmed milk. It is known as one of the most common intrinsically unstructured proteins, IUP's. In mammalian milk and its products it occurs as micelles made up of four major components [1.2] $\alpha_{s1}$-, $\alpha_{s2}$-, β- and κ-casein. The function of caseins is to store and transport bio-available metal ions (especially, Ca(II) and Mg(II)) by sequestering and transporting them from mother to the neonates [3-6].

Caseins possess a number of favorable characteristics suitable for the development of hydrogel biomaterials, such as high hydrophilicity, good biocompatibility particularly in oral delivery applications, lack of toxicity and availability of reactive sites for chemical modification. In aqueous solution single protein behaves as flexible, disordered, polyelectrolyte-like molecule [7], therefore, it is easily integrated into polyelectrolyte films [8,9]. Since one of the main features of casein is its ability to bind calcium ions, therefore, it can be used in biotechnology and in biomedical applications to promote biomineralization. A good example are materials for dental implants, which surface modified by films containing calcium crystallites provide better osteointegration [10]. Due to unique physicochemical properties as natural polymeric surfactant caseins are good candidates for the preparation of conventional and novel drug delivery systems [11]. They can also serve as shield against radiation, particularly UV light, utilizing its strong absorbance around 200–300 nm [12,13]. Casein films have been shown to exhibit high tensile strength making them suitable as coatings for tablets [13-15]. Films containing caseins are used as water based paints or adhesives for labeling of glass containers [16]. Materials covered with casein containing films can be applied in diary industry for the prevention of calcium deposit formation. Caseins can be also used in biosorbents for removal of multivalent metal ions (zinc, cadmium, mercury, chromium) [17,18].

---

*Corresponding author: Lilianna Szyk-Warszynska, E-mail:ncszyk@cyf-kr.edu.pl

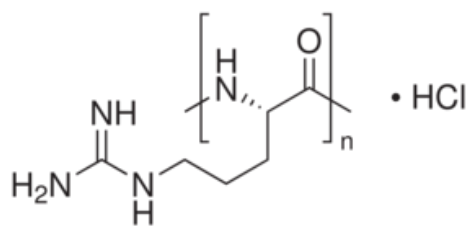

**Figure 1**. Poly L-arginine hydrochloride

Poly-L-arginine hydrochloride (PLArg) (Fig.1) is the positively charged, synthetic polyamino acid having one HCl per arginine unit. It is a biocompatible crystalline solid soluble in water. Poly-L-arginine was used as a component of microcapsules for protein or anticancer drug carriers [19]. It can serve as the immunostimulant in the anticancer vaccine [20]. Poly-L-arginine hydrochloride applications additionally include the polyelectrolyte film formation by the layer-by-layer (LbL) deposition technique [21] and the complexation with nucleic acids as sensor components [22] or for gene expression [23].

Sequential adsorption of charged nanoobjects at interfaces is a very versatile technique to form nanostructured thin films. In particular the sequential adsorption of polyelectrolytes (frequently referred to as the layer-by-layer (LbL) technique), which was introduced by Decher et al. [24, 25], has attracted much attention in the recent years. Due to the ability of producing multilayers films with well defined thickness and surface properties, the LbL technique can be useful in wide range of applications [26-29]. Embedding of proteins or other bio-active nanoparticles in polyelectrolyte multilayer films can contribute to the formation of surface nanostructures, which can be used in the biomaterial area [10, 29, 30].

In our previous work we demonstrated that using poly-L-lysine (PLL) as polycation the polyelectrolyte films containing casein can be formed at surface of silicon wafers [8]. We found that the films were stable in neutral and weakly basic conditions and in the NaCl concentration range from 0.015 to 0.15 M. Casein embedded in such films preserved its ability to bind calcium ions when exposed to calcium chloride solutions [8]. In the present paper we concentrated on formation of polyelectrolyte multilayer films containing α- and β-casein with another biocompatible cationic poly amino acid - poly-L-arginine. Films were formed at the surface of silicon wafers and quartz crystal plates covered with gold (QCM sensors) and their thickness/mass were analysed in both dry (ellipsometry) and wet (QCM-D) conditions. They were subjected to the post-treatment with the solutions of various composition to determine their stability. Films were also exposed to solutions containing calcium ions to evaluate the ability of their binding by casein embedded in the multilayers.

## 2. Experimental Setup

The α- casein (Cat. No. C6780-1G, min 70%, α-cas) and β-casein (Cat.No.C6905-1G, min 90%, β-cas) from bovine milk, poly-L-arginine hydrochloride (PLArg), MW 15kDa-70 kDa, (Cat.No. P7762), polyethyleneimine (PEI) MW 750 kDa, HEPES (N-2-Hydroxyethylpiperazine-N'-2- ethanesulfonic acid, Cat. No. H6147) and calcium chloride were obtained from Sigma. Sodium chloride pure p.a., hydrochloric acid, sodium hydroxide, sulfuric acid and hydrogen peroxide were obtained from POCh, Poland. Polished silicon wafers were purchased from On Semiconductor Czech Republic, a.s. (Cz/100T-0.5mm/(100)/P Type). Gold covered quartz crystals QCX301 for QCM-D experiments were obtained from Q-Sense, Sweden .

The substrates for the multilayer deposition, namely silicon wafers and QCM sensors were washed in piranha solution ($H_2SO_4:H_2O_2$,1:1), boiled three times in distilled water and rinsed with excess of distilled water. In such a way on the top of silicon wafers a layer of silica layer was formed. Its thickness was determined by ellipsometry multi-angle analysis and was equal to $3 \pm 1$ nm. To form the polycation layer 0.1 g/dm$^3$ PLArg solution in HEPES (10mM, 0.15 M NaCl, pH7.4) was used. Negatively charged α- or β-casein layers were adsorbed from 0.1 g/dm$^3$ solution of HEPES buffer (10 mM, 0.15 M NaCl, pH = 7.4). If not noted otherwise the washing steps were performed using HEPES buffer. Further details of the multilayer film formation method can be found elsewhere [8, 9]. Before the ellipsometric measurements films were carefully dried in the stream of argon.

The NanoZS Malvern dynamic light scattering (DLS) analyzer was used for zeta potential determination of α- and β-casein and poly-L-arginine. They were measured in 0.15 M NaCl and in dependence of pH of the solution, which was regulated by addition of HCl and NaOH. The Smoluchowski formula was used to calculate zeta potential from the electrophoretic mobility data.

The thickness and optical properties (refractive index, absorbance) of (PLArg/casein)$_n$ multilayer films adsorbed on silicon wafers Si/SiO$_2$ were determined by the single wavelength ($\lambda$ = 632 nm) imaging ellipsometer EP$^3$(Nanofilm) by fitting the "constant n.k" optical model for film consisting of two layers on solid support– (Si/SiO$_2$/PE film) to the measured values of the ellipsometric angles $\Psi$ and $\Delta$ [31]. The refractive index of PLArg/casein film was found by the multiangle analysis as n = 1.6, while its absorption coefficient k was negligible.

QCM-D (Q-Sense AB Gothenburg, Sweden) technique was used to analyze the mass, thickness and viscoelastic property of polyelectrolyte-casein films adsorbed on quartz, gold coated sensors. During the build-up of alternating layers of PLArg and casein one observed the decrease in frequency of crystal resonance and increase of dissipation of energy of oscillations, respectively. Changes in oscillating resonance frequency, $\Delta f$, are related to mass changes; a negative frequency shift is induced by the mass increase due to polyelectrolyte adsorption. The energy dissipation, $\Delta D$, is related to the viscous losses in the film. A high value of $\Delta D$ and its spread for the oscillation overtones indicates a soft film structure. The experimental $\Delta f$ and $\Delta D$ values for the multilayer films during their build-up and conditioning were interpreted basing on the viscoelastic Voigt model [32] and the thickness of the "wet" film was calculated as the best fit parameter.

The X-Ray Photoelectron Spectroscopy was utilized for the characterization of surface composition of silicon plates covered with multilayer (PLArg/cas)$_n$ films. XPS/ESCA measurements were carried out using a X-ray photoelectron spectrometer equipped with a semispherical analyzer SES R4000 (Gammadata Scienta) and monochromatic Mg K$\alpha$ source (1253.6 eV), 400 W power. The extraction angle of a photoelectron was fixed at 90˚ and the base pressure in the vacuum chamber during measurements was around 8x10$^{-10}$ mbar. Spectra were collected from area of the 3mm$^2$ and were worked out by CasaXPS 2.3.12 software. Background was approximated by Shirleys' algorithm. Lines of the spectra were deconvoluted (decomposed) and fitted by Voigt's function. For XPS experiments the multilayer films were prepared on silicon wafers using the same conditions as for ellipsometry. Further details of the procedure of measurements are given elsewhere [8].

## 3. Experimental Results

In our previous work [9] we demonstrated that the optimal conditions for the formation of the multilayer PLL/casein films at pH 7 concerning electrolyte concentration is 0.15 M NaCl i.e. close to the physiological conditions. As it is illustrated in Figure 2 in those conditions zeta potential of caseins is negative between –40 and –60 mV, while the zeta potential for poly-L-arginine is positive, +30 mV, that suggests that the same conditions are favorable for formation of the multilayer films with PLArg.

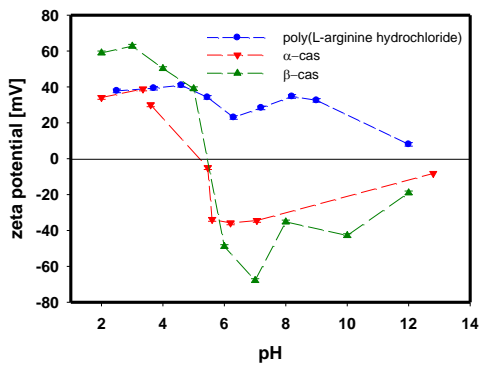

**Figure 2.** The dependence of zeta potential of $\alpha$- and $\beta$-casein and poly-L-arginine on pH of their solutions. Points represent experimental data. Lines were drown to guide the eye.

In the Figure 3 the dependence of the ellipsometric thickness of multilayers formed from the solution of poly-L-arginine and casein with concentrations of 0.1 g/dm$^3$ dissolved in HEPES (HEPES=10mM, NaCl=0.15M, pH=7.4) is shown.

One can see that the thickness of multilayer film increases with the number of adsorbed layers. It can be also observed that multilayer films with $\alpha$-casein are thicker than those with $\beta$-casein, which is in contrast to the higher surface activity of $\beta$-casein [33]. For instance (PLArg/$\alpha$-cas)$_4$ film is equal to 45±0.1 nm and

(PLArg/β-cas)$_4$ film has only 29±0.1 nm thickness. That phenomenon was observed before for (PLL/casein) films [8,9]

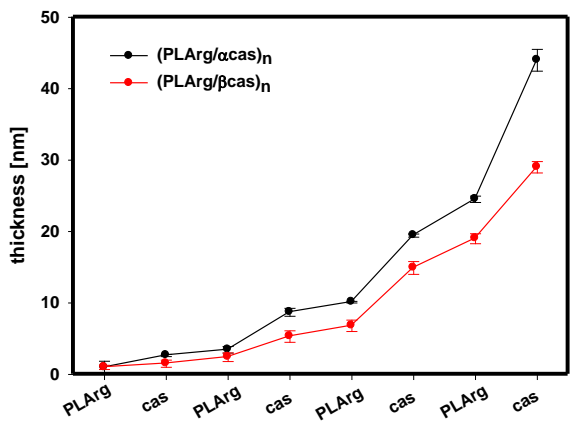

**Figure 3**. Ellipsometric thickness of (PLArg/cas) multilayer films formed on Si/SiO$_2$ plates after deposition of consecutive layers. Concentration of PLArg and casein solution $c_{PLArg,cas}$=0.1 g/dm$^3$ of HEPES ($c_{HEPES}$=10mM, I=0.15M NaCl, pH 7.4). Points represent experimental data. Lines were drown to guide the eye.

and can be explained by the block copolymer structure of α-casein [5] and/or by the differences in molecular size of caseins (molecular size of α-cas is around 9 nm while β-cas - 7 nm [34-36]). Both (PLL/casein)$_n$ and (PLArg/casein)$_n$ in 0.15 M NaCl exhibit fast growth of the film thickness with the number of layers, which suggests strong polyelectrolyte interpenetration [37] or/and formation of aggregates during the deposition process. The observed much larger increment of the film thickness after deposition of the casein than of the PLArg layer is caused by the difference in the cross-section of casein (7 – 9 nm) and PLArg molecule (1 – 2 nm) and most probably also by the deposition of casein aggregates during film formation causing irregular grow of the film. Increase of film thickness for the first 3-4 layer is much smaller than for one observed for next five layers. It is a common feature of polyelectrolyte films connected with the build-up of the first precursor layers of the film, which are adjacent to the silicon surface [38, 39]. The effect of the slower growth in the precursor layer can be circumvented by application of the anchoring layer of branched polyethyleneimine (PEI) [21, 39]. In Figures 4A and B the effect of the anchoring layer of PEI on the growth of the films containing α- and β-casein is shown, respectively. One can observe a big difference in the film thickness adsorbed on the bare silicon wafers and wafers where the first layer was branched PEI acting as the anchoring network for formation of the consecutive layers. Their growth is uniform from the very first layers, which results in much thicker films.

Formation of (PLArg/cas)$_n$ and PEI(cas/PLArg)$_n$ multilayer films from solutions having the same physicochemical parameters as those used for preparation of polyelectrolytes films on Si/SiO2 plates were also analyzed by QCM-D. The experimental data were collected in situ during sequential adsorption of polyelectrolytes on QCM-D sensors. It means that the films mass and thickness determined by quartz microbalance concern wet films as opposed to the ellipsometric data obtained for films adsorbed on silicon wafers and dried before measurements. The comparison of those data is given in Figures 5 and 6. The dependence of the "wet" film thickness on the number of deposited layer exhibited the same trends as observed for "dry" films, i.e., the PLArg/α-casein films were thicker than respective PLArg/β-casein ones and ones build on the PEI anchoring layer were thicker than films without that layer. However, we can observe that films measured by QCM-D were swollen, almost two times thicker than those measured by ellipsometry (see Figure 4A,B and 5A,B) due to large content of buffer penetrating multilayer film adsorbed on the QCM sensors. Hydration of the polyelectrolytes favors extended conformation of their chains, which form sponge-like film. That was also supported by a high value of the dissipation, increasing with the film thickness.

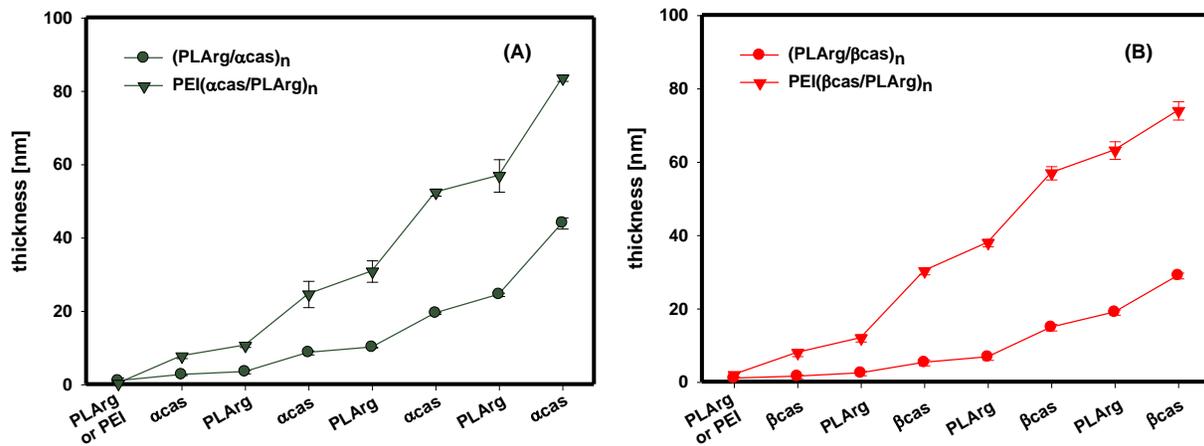

**Figure 4.** Ellipsometric thickness of (PLArg/cas)n and PEI(cas/PLArg)n multilayer films formed on Si/SiO2 plate after deposition of consecutive layers: (A) for α-casein, (B) -for β-casein; the condition for films deposition as in Fig.3.

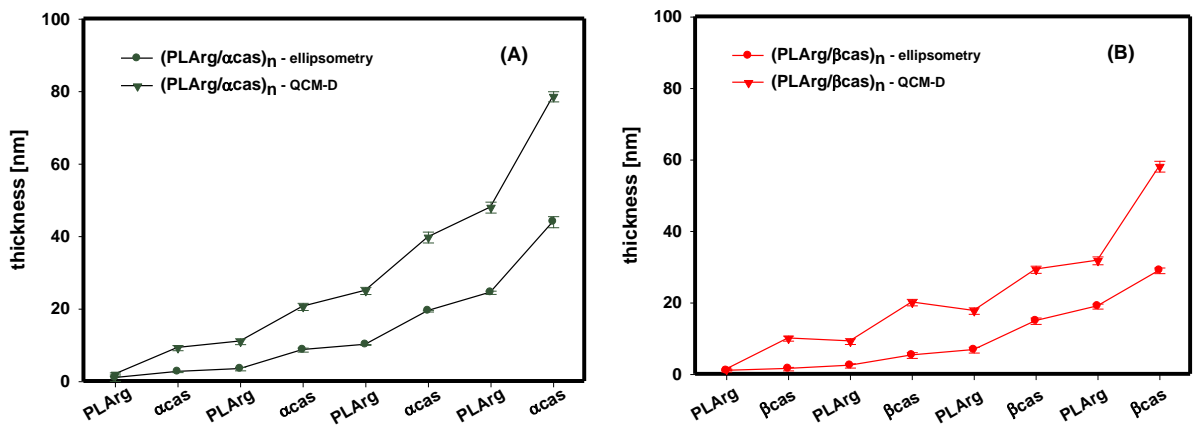

**Figure 5.** The comparison of the ellipsometric "dry" with the QCM-D determined "wet" thickness of (PLArg/cas)$_n$ multilayer films; (A) – for α-casein, (B) -for β-casein. Concentration of PLArg and casein solution $c_{PLArg,cas}=0.1$ g/dm$^3$ of HEPES ($c_{HEPES}$=10mM, I=0.15M NaCl, pH 7.4). Points represent experimental data. Lines were drown to guide the eye.

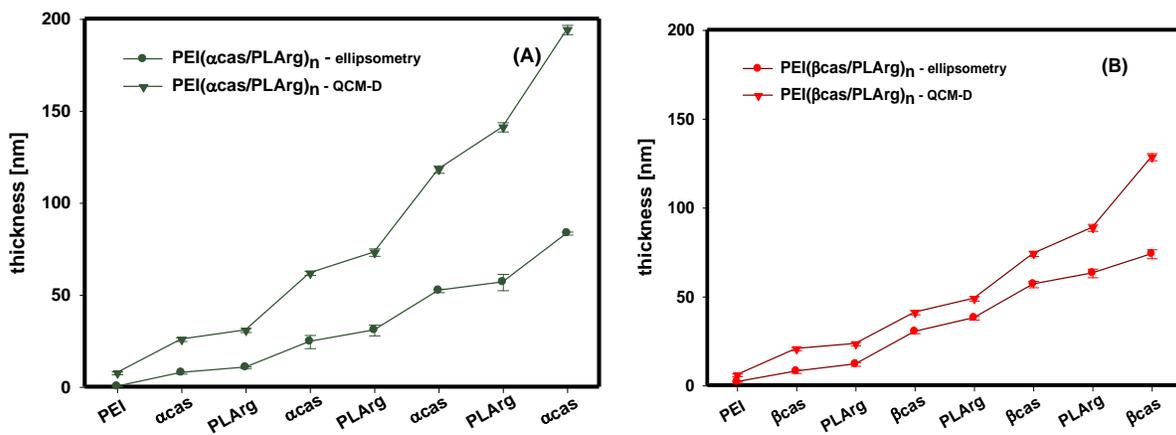

**Figure 6.** The comparison of the ellipsometric "dry" with the QCM-D determined "wet" thickness of PEI(cas/PLArg)$_n$ multilayer films; A – for α-casein, B -for β-casein. Concentration of PLArg and casein solution

$c_{PLArg,cas}$=0.1 g/dm$^3$ of HEPES ($c_{HEPES}$=10mM, I=0.15M NaCl, pH 7.4). Points represent experimental data. Lines were drown to guide the eye.

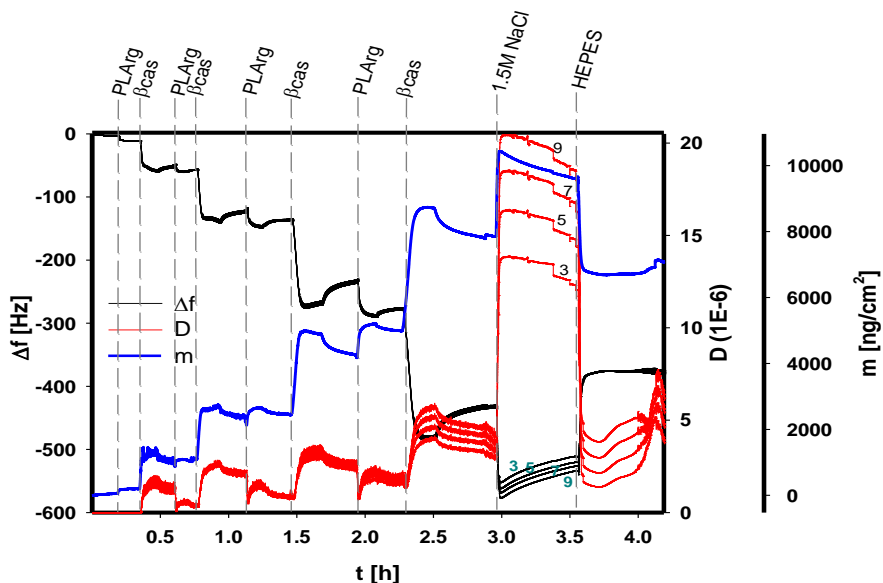

**Figure 7.** The example of QCM-D measurement of frequency shift and dissipation increase during adsorption of (PLArg/β-cas)$_4$ multilayer film and its post-treatment (in 1.5 M NaCl); red and black lines represent 3, 5, 7 and 9 QCM crystal oscillation overtone (red lines- dissipation, black lines – frequency); blue line represent mass of the adsorbed film, which was calculated using the viscoelastic Voigt model. The gray dashed perpendicular lines mark times of injection of the given solution. Concentration of PLArg and casein solution $c_{PLArg,cas}$=0.1 g/dm$^3$ of HEPES ($c_{HEPES}$=10mM, I=0.15M NaCl, pH 7.4).

To determine the stability of casein containing films we formed them *in situ* at the surface of QCM crystal and then exposed to the solutions of various pH and composition. Changes of the films adsorbed mass and energy dissipation during film formation and conditioning were measured by QCM-D and the mass of the film was calculated by Voigt viscoelastic model. The example of a single experimental run for the (PLArg/β-cas)$_4$ is illustrated in Figure 7. The multilayer film was deposited in HEPES buffer at I = 0.15 M NaCl and pH 7.4. One can see the systematic decrease of the crystal oscillation frequency (and corresponding increase of the adsorbed mass) with the deposition of the consecutive layers, the larger one associated with adsorption of casein. The dissipation of the crystal oscillation simultaneously increases. At 3h after forming the film it was exposed to the solution with higher ionic strength, I = 1.5 M NaCl, for 30 minutes and finally rinsed with HEPES (10mM, 0.15M NaCl, pH7.4). When (PLArg/βcas)$_4$ film was treated with the solution of higher ionic strength we observed increase of adsorbed mass and dissipation associated with the large split of the signals from five overtones. That can be interpreted as swelling of the multilayer due to the high ionic strength [40,41] and increased water content in the film. The dissipation and separation of the signals from five overtones strongly decreases when HEPES solution is introduced back into the QCM-D cell indicating that the film de-swelled but smaller frequency shift when compared to one before the treatment, evidences that some of the adsorbed layer was removed from the surface.

Analogous experiments were performed for QCM substrates with (PLArg/α- or β-casein)$_n$ films exposed after their formation to solutions of sodium chloride with the concentrations (0.15M, 0.75M, 1.5M), solutions with I = 0.15 M NaCl and adjusted pH values (2, 4, 7.4, 9 and 11) and mixture of sodium and calcium chloride ($c_{CaCl2}$ = 0.15 M, 0.15 M NaCl HEPES -10 mM). The results of these experiments are presented in Figures 8 - 10 illustrating changes of mass of multilayer films during deposition and conditioning steps. As it can be seen in Figures 8A and 8B, the films containing α- or β-casein are fairy stable when exposed to the osmotic stress induced by higher concentration of sodium chloride. We can observe only a minor decrease of the mass of the multilayer films, not exceeding 15%, when they were treated with the 0.75 M or 1.5M NaCl solutions. In Figure 9 the changes of mass during the formation of (PLArg/casein)$_4$ films and after their post-treatment in the solutions with I = 0.15 NaCl and pH value ranging from 2 to 11, are shown. It can be seen that in strongly acidic conditions (pH 2) films are not stable and are almost totally removed from the surface of QCM crystal. That is due to the reversal of charge of casein from negative to positive one at low pH (see Fig. 2). Therefore, in pH 2 there are no electrostatic interactions keeping together PLArg and casein in the film. Note that the gold surface bears negative charge in the whole pH range [42], therefore, some leftovers of the film, predominantly PLArg, can remain at the surface. In the strongly basic conditions (pH 11) the removal only of the last layer from the multilayer films is observed. In those conditions the zeta potential of poly-L-arginine (see Figure 2) is around +15 mV and for α-casein and β-casein is c.a. -20 mV and -30 mV, respectively, therefore, the internal layers are intermixed and stable. The observed relative stability of PLArg/cas films at high pH is in contrast with much lower stability observed previously for PLL/cas films in those conditions [9]. In weakly acid condition of the post-treatment (pH 4) the α-casein containing films seem to be more stable than the ones with β-casein. Although at these conditions protein molecule possesses net positive charge, it contains negative regions and that combining with hydrophobic interactions [43] contributes to stability of these films. Similarly at weakly basic conditions, pH 9, when both caseins are negatively charged the stability of the α-casein containing films seems to be slightly higher. In the neutral conditions, films containing both types of casein are stable that is an important factor for the potential biomedical applications of casein containing films for surface modification of materials.

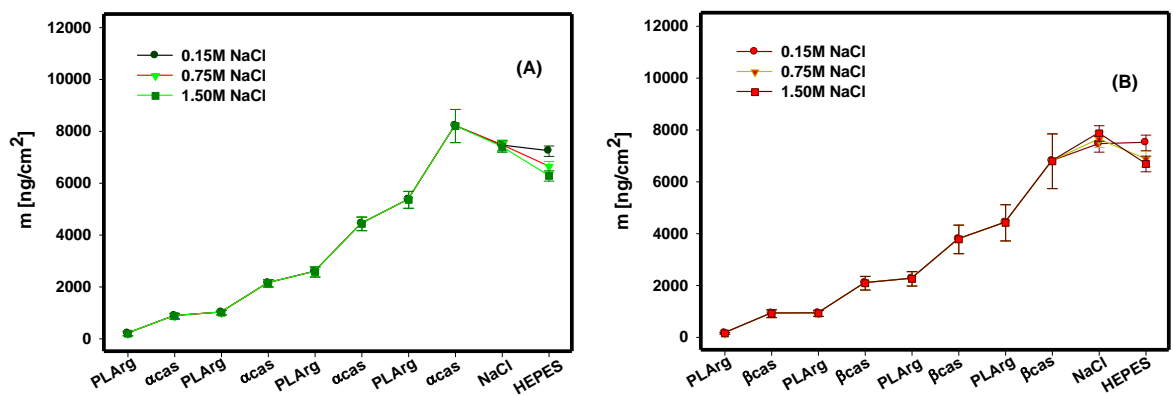

**Figure 8**. Changes of mass of (PLArg/casein)$_4$ film during its deposition at the surface of QCM sensors and post-treatment in NaCl solution with various concentration; (A) – for α-casein, (B) -for β-casein. Concentration of PLArg and casein solution $c_{PLArg,cas}$=0.1 g/dm$^3$ of HEPES ($c_{HEPES}$=10mM, I=0.15M NaCl, pH 7.4). Points represent experimental data. Lines were drown to guide the eye.

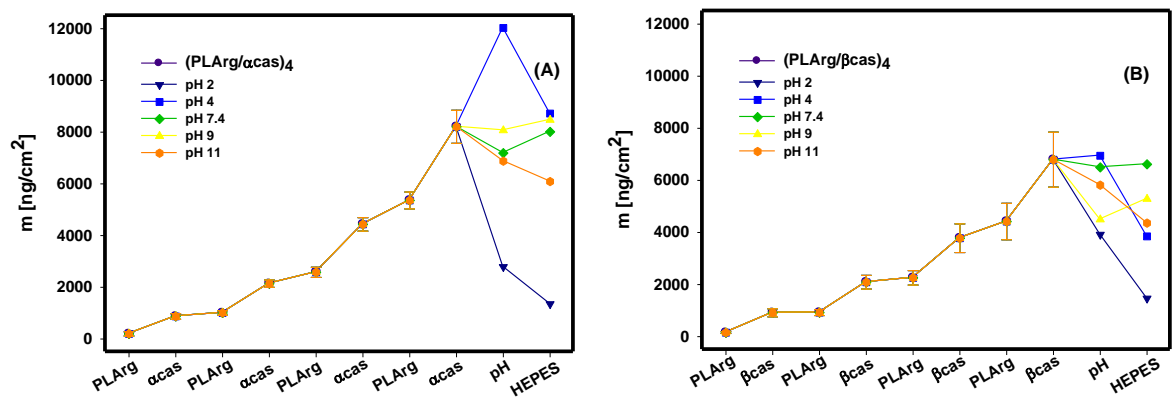

**Figure 9.** Changes of mass of (PLArg/casein)$_4$ film during its deposition at the surface of QCM sensors and post-treatment in 0.15 M NaCl solution with various pH; (A) – for α-casein, (B) -for β-casein.

Multilayer films with the composition (PLArg/cas)$_4$, (PLArg/cas)$_3$ and (PLArg/cas)$_2$ formed at the surface of QCM sensors were selected for the investigation of their stability in HEPES, 0.15 M NaCl, and 0.15 M CaCl$_2$ mixture. The results are shown in Figure 10. As it is illustrated in the Figure the almost rigid structure of the films, after being treated with CaCl$_2$ slightly swells due to the increase of the ionic strength of the environment, but when at the end of the experiment the surface with the film is rinsed with HEPES (10mM, 0.15 M NaCl, pH7.4) solution, the mass of the film decreases. The relative decrease is higher for films containing β-casein and can be interpreted as removal of the top film layer, destabilized by the Ca ions treatment.

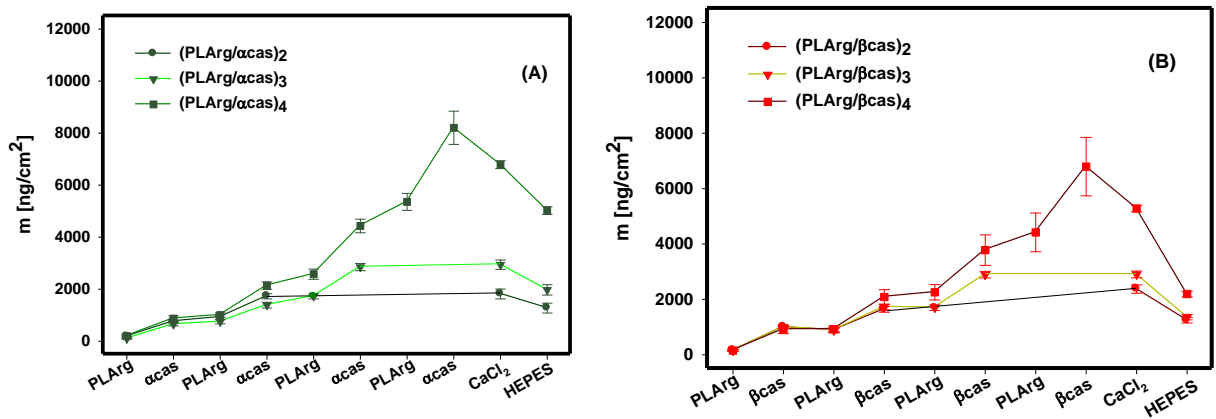

**Figure 10.** Changes of mass of (PLArg/casein)$_4$ film during its deposition at the surface of gold QCM sensors and post-treatment in 0.15 M CaCl$_2$ solution in HEPES with 0.15 M NaCl; (A) – for α-casein, (B) -for β-casein.

To verify the hypothesis whether, despite of the mass loss during the post-treatment, some calcium is retained in the film by casein sequestration, we performed the XPS elemental analysis for (PLArg/cas)$_4$ before and after contact with the HEPES (10mM, 0.15 M NaCl, pH 7.4) solution containing 2mM and 50 mM CaCl$_2$ for 24 hours. Before the analysis the samples were soaked for 4 h in the distilled water to avoid salts crystallization. The thickness of the films measured by ellipsometry before and after 24 hour post-treatment was compared. No significant changes was observed (data not shown). The Ca 2p line in the XPS spectrum could be detected in the (PLArg/α-cas)$_4$ sample contacted with calcium ions, while for the untreated sample this line was in the range of noise. We found the calcium content of 0.15-0.2 at % in the multilayer films regardless of Ca$^{2+}$ concentration in the conditioning solution, which seems to be slightly higher than for (PLL/casein) films [8]. It most probably means that the saturation level of calcium in the multilayer was already attained during treatment with 2 mM CaCl$_2$ for 24 hours. Calcium was bound to the casein since no chlorine was detected in the XPS spectrum. The results for (PLArg/β-cas)$_4$ were ambiguous but suggesting lower ability of this casein to bind calcium in the multilayer film, however that difference between alfa- and beta-casein was observed before in the case of PLL/casein multilayers [8].

## 4. Conclusions

The sequential adsorption of polyelectrolytes and proteins is a suitable technique for formation of surface nanostructures, which can be used in a biomaterial area. We investigated formation of polyelectrolyte multilayer films containing α- and β-casein together with biocompatible synthetic polyaminoacid poly-L-arginine. Using ellipsometry and QCM-D measurements we determined thickness and adsorbed mass of multilayer films containing alfa- and beta-casein and their stability in the conditions of various ionic strength and pH. We observed that multilayer films with α-casein were thicker than those with β-casein due to the copolymer structure of α-casein and its larger molecular size. Similarly as for the purely polyelectrolyte multilayers, application of high molecular weight branched polyethyleneimine anchoring layer induces more uniform growth of the films containing casein and that leads to thicker films. Comparison of the "dry" film thickness (measured by ellipsometry) with one measured in aqueous solution by QCM-D suggests up to 100% swelling of the films and strong hydration. Films deposited were not resistant to solutions of pH 2 due to the charge reversal of casein. At pH 11, despite low positive charge of PLArg only the external casein layer could be removed, while the intermixed internal layers remain stable. Similar behavior was observed at weakly acidic conditions, pH 4. Films remain stable in the neutral and weakly basic conditions that includes HEPES buffer, which is widely used in cell culture and biomedical experiments. At the conditions of high ionic strength (1.5 M NaCl or 0.15 M NaCl with 0.15 M $CaCl_2$) films swell but their swelling is reversible. Analysis of the XPS spectra of the (PLArg/casein) films exposed to $CaCl_2$ solutions revealed that the calcium ions can be bound by the casein embedded in the multilayers. That makes them promising coatings, promoting biomineralization at the surface of materials used for medical applications, e.g. for implants manufacturing.

## Acknowledgments

This work was partially supported by the National Science Centre (NCN) project NN 204 546 639 the Marian Smoluchowski Krakow Research Consortium - a Leading National Research Centre KNOW supported by the Ministry of Science and Higher Education and the COST Action CM1101

## References

[1] Qi, P.X.; Wickham, E.D.; Farell Jr., H.M., Prot. J. 23 (2004) 389-402.
[2] Swaisgood, H.E., In: Fox PF, editor. Advanced Diary Chemistry Proteins, New York: Elsevier, (1992) p.63-110
[3] Sahu, A.; Kasoju, N.; Bora, U., Biomacromolecules, 9(10) (2008) 2905-2912.
[4] Smyth, E.;. Clegg, R.A; Holt, C., Int.J.Dairy Tech. 57 (2-3) (2004) 121-126.
[5] Horne, D.S., Curr. Opin. Colloid Interface Sci.7 (2002) 456-461.
[6] Tompa, P., Trends Biochem. Sci., 27 (2002) 527-533.
[7] Maheshwari, R.K.;. Singh, A.K.; Gaddipati, J.; Srimal, R.C., Life Sci. 78 (2006) 2081-2087.
[8] Szyk-Warszyńska, L.; Gergely, C.; Jarek, E.; Cuisinier, F.; Socha, R.P.; Warszyński, P., Colloids Surf., A, 343 (2009) 118–126.
[9] Szyk-Warszyńska, L.; Piekoszewska, J.; Warszyński, P, Adsorption 16 (2010) 241–248.
[10] Furedi-Milhofer, H.; Sikiric, M.; Yosef, P.B.; Cuisinier, F.J.G.; Gergely, C., US patent application #60-428.72 (2004).
[11] Livney, Y.D., Curr. Opin. Colloid Interface Sci. 15 (2010) 73–83.
[12] Swaisgood, H.E., P.F. Fox, P.L.H. McSweeney (Eds.), Advanced Dairy Chemistry. Vol. 1: Proteins Part A, Kluwer Academic/Plenum Publishers, New York, (2003) 139–202.
[13] Semo, E.; Kesselman, E.; Danino, D.; Livney, Y.D., Food Hydrocoll. 21 (2007) 936–942.
[14] Motoki, M.; Aso, H.; Seguro, K.; Nio, N., Agric. Biol. Chem. 51 (1987) 993–996.
[15] Abu Diak, O.; Bani-Jaber, A.; Amro, B.; Jones, D.; Andrews, G.P., Trans. IChemE, C, Food Bioproducts Process. 85 (2007) 284–290.
[16] Muller-Buschbaum, P.; Gebhardt, R.; Maurer, E.; Bauer, E.; Gehrke, R.; Doster, W., Biomacromolecules 7 (2006) 1773-1789.
[17] Mishra, S.P.; Tiwari, D.; Dubey, R.S.; Mishra, M., Bioresour. Technol. 63 (1998) 1–5.
[18] Seki, H.; Suzuki., A., J. Colloid Interface Sci. 249 (2002) 295–300.
[19] Lan, Q.; Wang, Y.; Wang, S.; Liu, Y., IFMBE Proceedings 19 (2008) 32-35.
[20] Mattner, F.; Fleitmann, J.-K.; Lingnau, K.; Schmidt, W.; Egyed, A.; Fritz, J.; Zauner, W.; Wittmann, B.; Gorny, I.; Berger, M.;Kirlappos, H.; Otava, A.; Birnstiel, M. L.; Buschle; M., Cancer Res March 1,62 (2002) 1477.
[21] Elzbieciak-Wodka, M.; Warszynski, P., Electrochimica Acta, 104 (2013) 348-357.
[22] Sukhorukov, G.B.; Montrel, M.M.; Petrov, A.; Shabarchina L.I.; Sukhorukov, B.I., Solid-State Sensors and Actuators, 1995 and Eurosensors IX.. Transducers '95. The 8th International Conference on (Volume:2 ) 524-527.
[23] Kim, E.-J.; Shim, G.; Kim, K.; Kwon, I. Ch.; Oh, Y.-K.; Shim, Ch.-K., J. GENE MED. 11(9) (2009) 791-803.
[24] Decher, G.; Hong, J.-D., Ber. Bunsen.-Ges. Phys.Chem. 95 (1991) 1430.
[25] Decher, G., Science 277 1 (1997) 232.
[26] Castelnovo, M.; Joanny, J–F., Langmuir 16 (2000) 7524.
[27] Hammond, P.T., Curr. Opin. Colloid Interface Sci. 4 (2000) 430.
[28] Schönhoff, M., Curr. Opin. Colloid Interface Sci. 8 (2003) 86.

[29] Decher, G.; Schlenoff J.B.; Eds., Multilayer thin films: sequential assembly of nanocomposite materials, 2$^{nd}$ Edition, Wiley-VCH 2012.
[30] Tang, Z.; Wang, Y.; Podsiadlo, P.; Kotov, N. A., Adv. Mater 18 (2006) 3203–3224.
[31] Keddie, J.L., Curr. Opin. Colloid Interface Sci. 6 (2001) 102-110.
[32] Voinova, M.V.; Rodahl, M.; Jonson, M.; Kasemo, B., Physica Scripta 59 (1999) 391-396.
[33] Dickinson, E., Rolfe, S.E., Dalgleish, D.G., Food Hydrocoll, 2 (1988) 187.
[34] Thomas F. Kumosinski & Michael N. Liebman, Eds. ACS Symposium Series 576, American Chemical Society, Washington, DC 1994, pp 368-390.
[35] Thomas F. Kumosinski, Eleanor M. Brown, and Harold M. Farrell, Jr., Predicted Energy-Minimized Alpha-S1-Casein Working Model in MOLECULAR MODELING: FROM VIRTUAL TOOLS TO REAL PROBLEMS,
[36] T. F. Kumosinski, E. M. Brown, and H. M. Farrell, Jr., Three-Dimensional Molecular Modeling of Bovine Caseins: An Energy-Minimized Beta-Casein Structure (1993) *Journal of Dairy Science*, 76:931-945.
[37] Porcel, C.; Lavalle, P.; Ball, V.; Decher, G.; Senger, B.; Voegel, J.-C.; Schaaf, P., Langmuir 22 (2006) 4376-4383.
[38] Buron, C.C.; Filiâtre, C.; Membrey, F.; Perrot, H.; Foissy, A., Journal of Colloid and Interface Sciences 296 (2006) 409-418.
[39] Kolasińska, M.; Krastev R.; Warszyński, P., J. Colloid Interface Sci. 305 (2007) 46-56.
[40] Dubas S.T, Schlenhoff J.B., Macromolecules, 32 (1999) 8153.
[41] Irigoyen, J., Han, L., Llarena, I., Mao, Z., Gao, C., Moya, S. E., Macromol. Rapid Commun. (2012) DOI: 10.1002/marc.201200471
[42] Park, S-E.; Park, M-Y.; Han, P-K.; Lee, S-W., Bull. Korean Chem. Soc. Vol. 27 No. 9 (2006) 1341-1345.
[43] Dickinson, E., Colloids Surf., A 288 (2006) 3–11.